\documentclass[draft]{agujournal2019}
\usepackage{url} 
\usepackage{lineno}
\usepackage{soul}
\draftfalse

\journalname{JGR: Machine Learning and Computation}

\begin{document}

%
%


\title{Machine learning potential for serpentines}

%
%











\authors{Hongjin Wang\affil{1}, Chenxing Luo\affil{1}, Renata M.~Wentzcovitch\affil{1,2,3,4}}

\affiliation{1}{Department of Applied Physics and Applied Mathematics, Columbia University, New York, New York 10027, USA}
\affiliation{2}{Department of Earth and Environmental Sciences, Columbia University, New York, New York 10027, USA}
\affiliation{3}{Lamont--Doherty Earth Observatory, Columbia University, Palisades, New York 10964, USA}
\affiliation{4}{Data Science Institute, Columbia University, New York, New York 10027, USA}

\correspondingauthor{Renata M.~Wentzcovitch}{rmw2150@columbia.edu}



\begin{keypoints}
\item We develop a deep learning-based machine learning potential for serpentine minerals based on the r$^2$SCAN meta-GGA DFT functional suitable for molecular dynamics simulations. 
\item We accurately reproduce experimental EoSs of brucite, lizardite, and antigorite minerals at 300~K under subduction zone pressures. 
\item We investigate all the naturally existent antigorite polysomes with the periodicity parameter $m$ = 13--24, shedding light on their relative stability, especially that of antigorite over lizardite at lower temperatures for $m > 21$.

\end{keypoints}

%
%

%
%


\begin{abstract}
Serpentines are layered hydrous magnesium silicates (MgO$\cdot$SiO$_2\cdot$H$_2$O) formed through serpentinization, a geochemical process that significantly alters the physical property of the mantle.
They are hard to investigate experimentally and computationally due to the complexity of natural serpentine samples and the large number of atoms in the unit cell.
We developed a machine learning (ML) potential for serpentine minerals based on density functional theory (DFT) calculation with the r$^2$SCAN meta-GGA functional for molecular dynamics simulation.
We illustrate the success of this ML potential model in reproducing the high-temperature equation of states of several hydrous phases under the Earth's subduction zone conditions, including brucite, lizardite, and antigorite.
In addition, we investigate the polymorphism of antigorite with periodicity $m$ = 13--24, which is believed to be all the naturally existent antigorite species.
We found that antigorite with $m$ larger than 21 appears more stable than lizardite at low temperatures.
This machine learning potential can be further applied to investigate more complex antigorite superstructures with multiple coexisting periodic waves.
\end{abstract}

\section*{Plain Language Summary}
Serpentines, including antigorites and lizardites, are layered hydrous minerals formed through serpentinization, a geochemical process that significantly alters the physical property of the mantle. 
They participate in many geodynamic processes, such as Earth's deep water cycle, and may induce Earthquakes. 
Despite their geophysical significance, their properties remain unknown due to experimental and computational challenges.
Here, we developed a machine learning potential model that allows us to perform accurate molecular dynamics simulations on serpentine minerals.
We demonstrate the success of this machine learning potential in the serpentine systems by reproducing the high-temperature equation of states, which are key parameters obtained from measurements.
We investigated all the naturally existent antigorite species and found that antigorite with $m$ in certain conditions appears more stable than lizardite at low temperatures.
%
%

%


%
%
%
%

\section{Introduction}

Serpentines are the product of serpentinization, a process where ultramafic rocks, particularly peridotites, react with water and transform into hydrous magnesium silicate minerals (MgO$\cdot$SiO$_2\cdot$H$_2$O), including lizardite, chrysotile, and antigorite.
Throughout the process, rock density reduces from $\sim$3.3~g/cm$^3$, typical of olivine (the main component of peridotites), to 2.6--2.7~g/cm$^3$, making fully serpentized rock as buoyant as the continental crust.
Serpentines play a crucial role at shallow depths, particularly in fault zones, where they facilitate fault zone weakening and aseismic creep \cite{reinenFrictionalBehaviorSerpentinite1991,hilairetHighPressureCreepSerpentine2007}. 
In subduction zones, the dehydration of serpentines leads to slab embrittlement, potentially triggering seismic activities \cite{cholletKineticsMechanismAntigorite2011,reynardSerpentineActiveSubduction2013}.
With $\sim$13~wt\% of water stored in hydroxyl groups, serpentines act as major water carriers in subduction zones at crust to upper mantle depths. They serve as a key carrier for water to enter the Earth's deep water cycle, enabling subsequent geodynamic process \cite{ohtaniHydrationDehydrationEarth2021}.
However, despite their geophysical importance, serpentine minerals' structures and physical properties remain poorly understood due to experimental and computational challenges.

While modern experimental techniques, such as laser-heated diamond-anvil cells and multi-anvil press apparatus, can access pressure-temperature ($P$-$T$) conditions far beyond those relevant to serpentine formation, the properties of serpentine remain poorly constrained. 
This is largely due to the complexity of natural serpentine rock samples, which often include variable mixtures of serpentine polytypes (lizardite, chrysotile, antigorite) along with other minerals (e.g., olivine, magnetite, talc, tremolite). These minerals, formed before or during serpentinization and metasomatism, occur in varying proportions, leading to significant variability in rock properties \cite{vitiDeformationProcessesTextural2018, shen2020changes, jiIdentifyingSerpentineMinerals2024}.
Even among serpentine polytypes, structural differences---such as the periodicity and stacking of silicon tetrahedral and magnesium octahedral sheets---can lead to significant variations in their mechanical properties despite their compositional similarities \cite{shen2020changes}.
As a result, bulk rock properties cannot be reliably generalized without addressing the behavior of individual components and their proportions. 
While techniques transmission electron microscopy (TEM) or Raman spectroscopy (due to the presence of hydrogen) \cite{andreaniMicrostructuralStudyCrackseal2004,petriglieri2015micro,shen2020changes} may decipher the rock mineralogy and its microstructure, systematic investigations of the individual mineral phases, particular for the serpentine polytypes, remains critical for understanding the overall behavior of serpentinized rocks and for further assessing the effects of serpentinization on rock properties.

Computational approaches, particularly those based on the density functional theory (DFT), have been applied to serpentine polytypes \cite{mookherjeeStructureElasticitySerpentine2009,tsuchiyaFirstprinciplesCalculationElastic2013,ghaderiRelativeStabilityContrasting2015}.
However, standard functionals, LDA and GGA, often fail to accurately describe hydrogen bonds (H-bonds) in systems like brucite [Mg(OH)$_2$] or $\delta$-AlOOH \cite{wangInitioStudyStability2024, luo2022ab}. 
Even for lizardite, the simplest polymorph of serpentine, previous studies identify significant discrepancies with experiments \cite{mookherjeeStructureElasticitySerpentine2009, dengElasticAnisotropyLizardite2022}.
A further challenge in modeling antigorite arises from its complex, wavy structure, characterized by the periodicity of silicon tetrahedra along a wavelength (denoted by the parameter $m$).
Following a chemical formula of Mg$_{3m-3}$Si$_{2m}$O$_{5m}$(OH)$_{4m-6}$, antigorite with large $m$ can contain over 400 atoms in its unit cell, exceeding the computational capabilities of typical DFT-based methods, which generally limited to 200--300 atoms due to their cubic scaling with the number of valence electrons.
The unit cell of antigorite with a large $m$ is challenging even on state-of-the-art high-performance computing (HPC) systems.
Moreover, in quasiharmonic approximation (QHA) based methods, capturing H-bond disordering and intermediate-range ordering (as governed by ice rules) requiring even larger supercell, further escalating the computational cost \cite{raugeiPressureInducedFrustrationDisorder1999, luo2022ab, luoCijPythonCode2021a,wangInitioStudyStability2024}.
Intrinsic anharmonicity, which is not captured by QHA \cite{zhangPhononQuasiparticlesAnharmonic2014, zhuang2021thermodynamic,wang2023pgm}, is also expected when serpentine approaches the high-temperature stability of $\sim$1,000~K.

Recent advances in machine learning-based molecular dynamics offer promising solutions to these computational challenges.
Machine learning potentials, parameterized using DFT's description of the energy and forces of the system, allow for more efficient modeling of large systems with more accurate DFT functionals.
In these models, the total energy of the system is expressed as the sum of local atomic contributions, which are functions of descriptors representing the local chemical environment. 
Various machine learning methods can be used to fit energy as a function of these descriptors \cite{behlerGeneralizedNeuralNetworkRepresentation2007,bartokGaussianApproximationPotentials2010,thompsonSpectralNeighborAnalysis2015,shapeevMomentTensorPotentials2016,artrithEfficientAccurateMachinelearning2017,NEURIPS2018_e2ad76f2}. 
The Deep Potential Smooth Edition (\textsc{DeepPot-SE}) model \cite{NEURIPS2018_e2ad76f2} implemented in \textsc{DeePMD-kit} \cite{wangDeePMDkitDeepLearning2018, zengDeePMDkitV2Software2023} uses a deep neural network as the mapping function and has reliably describes complex H-bearing systems under extreme pressures and temperatures \cite<e.g.,>[]{PhysRevLett.126.236001,luo2024elasticityacousticvelocitiesdeltaalooh,luoProbingStateHydrogen2024}. This model enables accurate large-scale molecular dynamics simulations necessary to address the behavior of serpentine under finite $P,T$, where hydrogen bonding and anharmonicity play significant roles.

In this study, we focus on the antigorite and lizardite polytypes of serpentine. We developed a machine learning potential based on the r$^2$SCAN-DFT functional to systematically investigate the structures and compressive behaviors of serpentine and relevant systems at $P,T$ conditions of subducting slabs. 
Specifically, we investigate the polymorphism of antigorite with all naturally occurring periodicities ($m$ = 13--24).
We calculate lizardite, brucite, and antigorite compression curves at 300 K and compare them with existing measurements.
We also explore the relative phase stability of these minerals using enthalpy calculations obtained with our machine-learning model.

This paper is organized as follows.
Section~\ref{sec:method} introduces the deep learning potential method and DFT calculation parameters.
We show results and compare them with measurements and previous calculations in Section~\ref{sec:result}.
Summary and conclusions are presented in Section~\ref{sec:conclusion}.

\section{Method}
\label{sec:method}

\subsection{ML molecular dynamics potential} 

The neural network potential in this work was developed based on the \textsc{DeepPot-SE} model \cite{NEURIPS2018_e2ad76f2} implemented in \textsc{DeePMD-kit} v2.1 \cite{wangDeePMDkitDeepLearning2018, zengDeePMDkitV2Software2023}.
Two-body embedding with coordinates of the neighboring atoms (\texttt{se\_e2\_a}) was used for the descriptor.
The embedding network shape is (25, 50, 100).
The fitting network shape is (240, 240, 240).
The cutoff radius is 6 {\AA}, and the smoothing parameter is 0.5 {\AA}.
The model was trained using the \textsc{Adam} \cite{kingmaAdamMethodStochastic2017} optimizer for $1\times10^6$ training steps, with the learning rate exponentially decaying from $1\times10^{-3}$ to $3.51\times10^{-8}$ throughout the training process.
The loss function $\mathcal{L}(p_e, p_f)$ is \cite{wangDeePMDkitDeepLearning2018}
\begin{equation}
    \mathcal{L}(p_e, p_f)=p_e|\Delta e|^2+\frac{p_f}{3N}|\Delta f_i|^2 \,,
\end{equation}
where $p_e$ decays linearly from 1.00 to 0.02, and $p_f$ increases linearly from $1\times10^0$ to $1\times10^3$ throughout the training process.

\subsection{Active learning scheme}

The DP-GEN concurrent learning scheme \cite{zhangDPGENConcurrentLearning2020} was employed to create the training data set and to generate the potential. 
We randomly selected 14 serpentine configurations, including lizardite and $m = 17$ antigorite, to generate the initial potentials and kickstart the DP-GEN training process. The configurations were selected from various pressures, and the atoms were randomly displaced from their equilibrium positions.
We trained 4 candidate DP potentials initialized with different random seeds. They were used to perform $NPT$ DPMD simulations at 64 $P,T$s ($0<P<15$~GPa and $0<T<900$~K) for a few thousand timesteps. After the simulations, the error estimator (model deviation), $\epsilon_t$, is calculated every 50~MD steps based on the force disagreement between the candidate DPs \cite{zhangActiveLearningUniformly2019, zhangDPGENConcurrentLearning2020}:
\begin{equation}
\epsilon_t = \max_i \sqrt{\left< \Vert F_{w,i} (\mathcal{R}_t) - \left< F_{w,i} (\mathcal{R}_t) \right> \Vert^2 \right>}\,,
\end{equation}
where $F_{w,i}(\mathcal{R}_t)$ denotes the force on the $i$-th atom predicted by the $w$-th potential for the $\mathcal{R}_t$ configuration. For a particular configuration, if $\epsilon_t$ satisfies $\epsilon_\mathrm{min} \le \epsilon_t \le \epsilon_\mathrm{max}$, the configuration is collected, labeled with DFT forces and total energy, and added into the training dataset;
if $\epsilon_t < \epsilon_\mathrm{min}$, the configuration is considered ``covered'' by the current training dataset; if $\epsilon_t > \epsilon_\mathrm{max}$ the configuration is discarded.
After a few iterations, almost no new configurations are collected according to this standard ($> 99$\% are ``accurate'' for a few iterations), and the DP-GEN process is then completed.
Serpentine structures with $m$ = 13--24 configurations and brucite were subsequently included during the training process.
In total, it takes 6 DP-GEN iterations to generate a potential reaching satisfactory accuracy requirement for DPMD to describe brucite and serpentines at $300 < T < 900$~K and $0 < P < 15$~GPa.
The composite of the dataset generated in the DP-GEN iteration is listed in the Supporting Information Table~S1 in detail.

\subsection{DFT parameters}

Training and testing data sets were generated using on \textit{ab initio} calculations performed with the \textsc{Vienna Ab initio Simulation Package} (VASP)~v6.3 \cite{kresseEfficientIterativeSchemes1996a}. The regularized form of the SCAN functional (r$^2$SCAN) \cite{furnessAccurateNumericallyEfficient2020a} meta-GGA functional with PAW basis sets was adopted.
The cutoff energy for the plane-wave-basis set was set to 900~eV.
The Brillouin zones of brucite, lizardite, and antigorite were sampled using $\Gamma$-centered $4\times 4\times 4$, $4\times4\times4$, and $1\times2\times2$ $k$-point mesh respectively to ensure the convergence of 1~meV/atom.
The convergence criteria were $10^{-5}$~eV for the total energy and $10^{-3}$~eV/{\AA} for the atomic force. 

\subsection{MD simulations}

For validation purposes, Born-Oppenheimer molecular dynamics (BOMD) was conducted on brucite's $2\times2\times2$ supercells (144 atoms), lizardite's $2\times2\times2$ supercells (120 atoms), and $m = 17$ antigorite's unit cells (291 atoms) with Nosé-Hoover thermo-baro-stat \cite{hoover1996kinetic} and a time step of 0.2~fs.
These simulations were performed at 300~K and 0~GPa.


For the high-temperature equation of states, we performed DPMD simulations with $4\times4\times4$ (1152 atoms), $4\times4\times4$ (960 atoms), and $2\times2\times2$ supercells (1752--3336 atoms) for brucite, lizardite, and antigorite with all periodicities, respectively. 
We performed $NPT$ simulations for 1~ns to equilibrate the cell shapes, with a timestep of 0.2~fs using the Nosé-Hoover thermo-baro-stat obeying modular invariance \cite{wentzcovitch1991invariant}.
The pair correlation function was computed using BOMD and DPMD trajectories.
The configurations used in BOMD are unitcells generated from DPMD after 1~ns DPMD simulation.

\subsection{Chemical environment analysis}
We used the Smooth Overlap of Atomic Positions (SOAP) descriptor \cite{bartok2013representing} implemented in the DScribe \cite{dscribe2} package to investigate the local chemical environment of brucite, lizardite, and antigorite.
The cutoff for local regions was set to 6 {{\AA}}.
The number of radial basis functions and maximum degree of spherical harmonics were set to 12 and 8, respectively.
The brucite, lizardite, and antigorite structures were sampled from DPMD trajectories at 300, 600, 900~K, and 0--15~GPa.
The high-dimensional representation of the chemical environment is further visualized using principal component analysis (PCA). 

\section{Results and Discussion}
\label{sec:result}

\subsection{Benchmarking the deep potential}
\begin{figure}[htbp]
    \centering
    \includegraphics[width=0.8\textwidth]{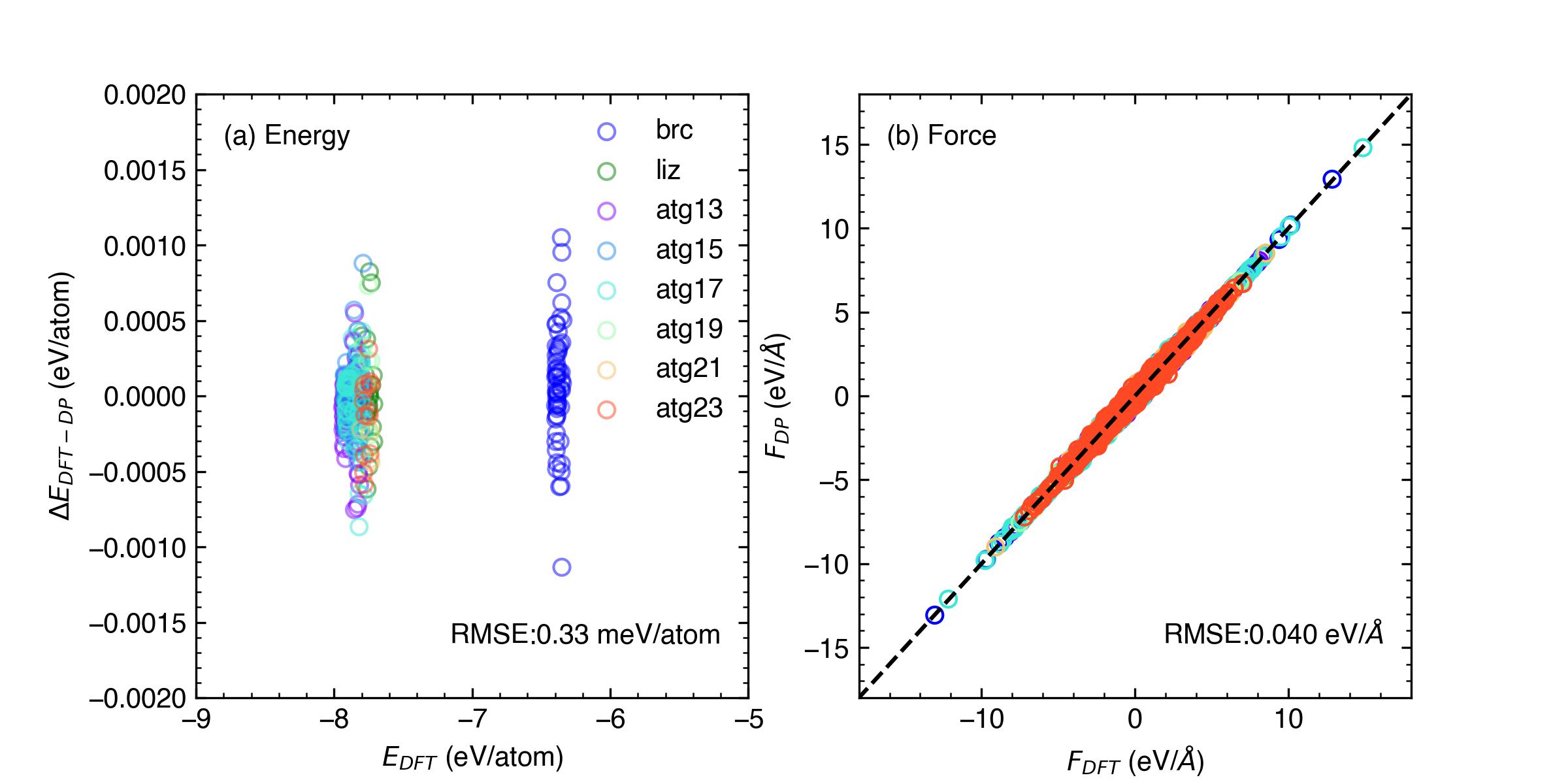}
    \caption{Comparison of energy and force from r$^2$SCAN-DFT and r$^2$SCAN-DP.}
    \label{fig:dp_test}
\end{figure}

In Fig.~\ref{fig:dp_test}, we compare the energy and force discrepancies between the r$^2$SCAN-DP's and r$^2$SCAN-DFT's predictions.
The absolute energy per atom formed two clusters, one for brucite and one for serpentines.
One is serpentine, including lizardite and antigorite with different periodicity, and the other one is brucite. 
The maximum difference in energy across all species is $\sim$1~meV/atom, with the RMSE of 0.33~meV/atom.
The force difference computed by r$^2$SCAN-DP vs.\ r$^2$SCAN-DFT is shown in Fig.~\ref{fig:dp_test}(b).
Overall, the force computed by r$^2$SCAN-DP is very close to r$^2$SCAN-DFT, with the RMSE value of 0.04~eV/{\AA}.

\begin{figure}[htbp]
    \centering
    \includegraphics[width=0.5\textwidth]{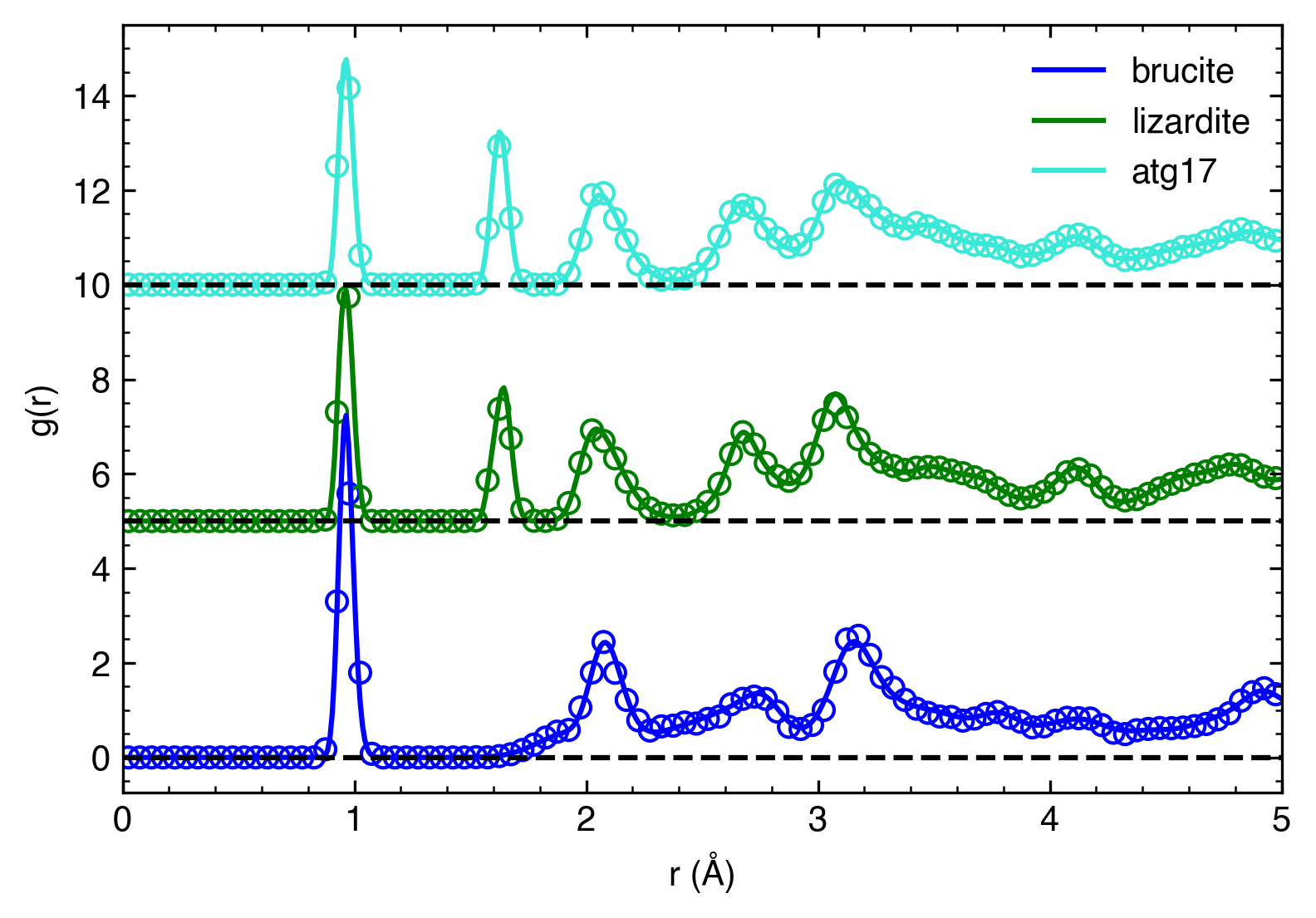}
    \caption{Comparison of the pair-correlation functions, $g(r)$, between r$^2$SCAN-DPMD and r$^2$SCAN-BOMD for brucite, lizardite, and antigorite $m=17$ at 0~GPa and 300~K.}
    \label{fig:dp_rdf}
\end{figure}

We also compare the structure described by r$^2$SCAN-BOMD and r$^2$SCAN-DPMD by comparing their pair correction functions, $g(r)$, as shown in Fig.~\ref{fig:dp_rdf}.
We compute the $g(r)$ for brucite, lizardite, and antigorite with $m=17$ using r$^2$SCAN-DPMD and r$^2$SCAN-BOMD.
The $g(r)$ of r$^2$SCAN-DPMD is almost indistinguishable from that of r$^2$SCAN-BOMD among all three systems,
indicating the accuracy of one machine learning potential model in describing these systems.

\subsection{Structures of antigorite with variable periodicity}

\begin{figure}[htbp]
    \centering
    \includegraphics[width=1\textwidth]{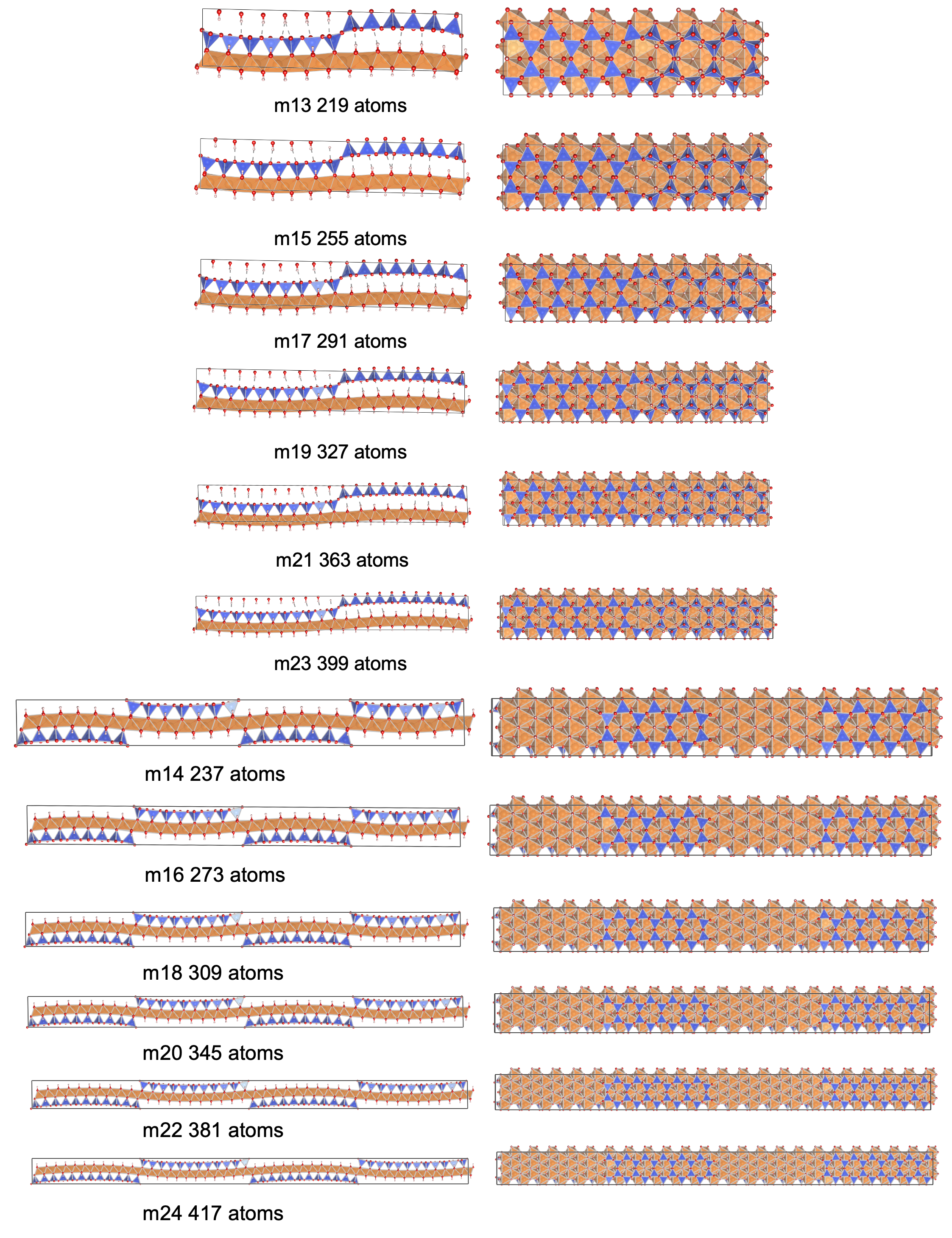}
    \caption{Top and side views of antigorite structures with different periodicity parameter $m$. 
    The black boxes indicate unit cells.
    The numbers of atoms in the captions are the numbers of atoms in the primitive cells.
    For antigorite with even $m$, the number of atoms in primitive cells is halved compared to conventional cells.}
    \label{fig:all_structures}
\end{figure}

Antigorite, with its chemical formula Mg$_{3m-3}$Si$_{2m}$O$_{5m}$(OH)$_{4m-6}$, has a wavy structure with 1-1 layered Si-O tetrahedral and Mg-O octahedral layers connected by H-bonds, very similar to lizardite and chrysotile.
The parameter $m$ is the number of tetrahedra along an entire wavelength (See Fig.~\ref{fig:all_structures}).
Antigorite exists in nature with various $m$ values, $13\leq m\leq 24$ \cite{melliniAntigoritePolysomatismBehaviour1987}.
When $m$ is odd, it has the space group $Pm$ \cite{capitaniModulatedCrystalStructure2004a};
when $m$ is even, it has the space group of $C2/m$ \cite{capitaniCrystalStructureSecond2006b}. 
Here, we generated structures of antigorite with $m$ from 13 to 24 following these symmetries for even and odd $m$ antigorite.
Fig.~\ref{fig:all_structures} shows top and side views of the structures of antigorite with $m$ = 13--24 \cite{capitaniModulatedCrystalStructure2004a,capitaniCrystalStructureSecond2006b}.
For antigorite with odd $m$, a long half wave of tetrahedra at the bottom alternates with a short half wave at the top, while for antigorite with even $m$, the structure is more symmetric with the same number of tetrahedra in both half waves.

\begin{figure}[htbp]
    \centering
    \includegraphics[width=.9\textwidth]{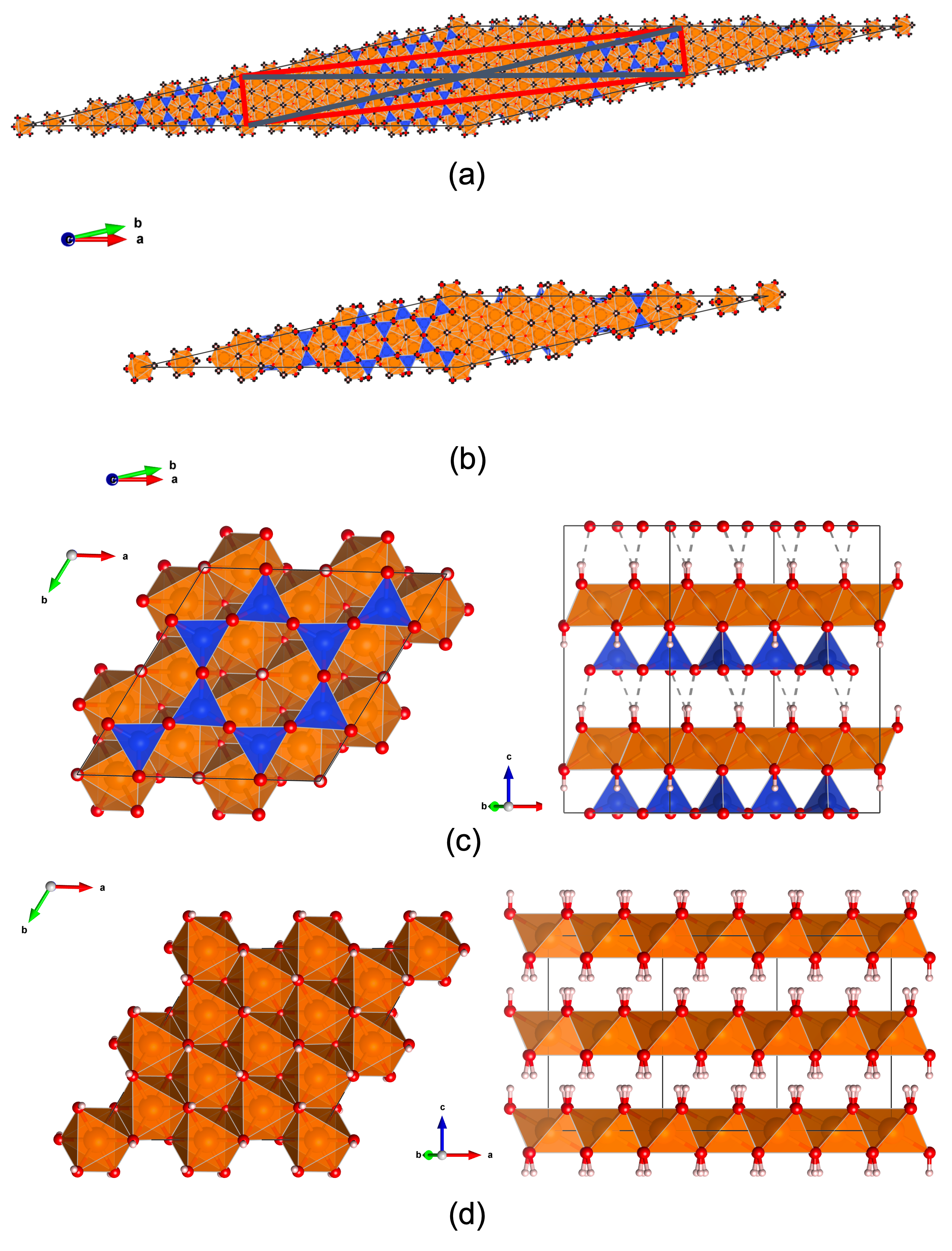}
    \caption{Relation of an $m=16$ antigorite primitive cell with the conventional unit cell as shown in Fig.~\ref{fig:all_structures}. 
    (a) Top view of a $2\times2\times1$ supercell of the primitive cell.
    The red rectangle shows a conventional unit cell.
    (b) Top view of an $m=16$ antigorite primitive cell.
    (c) Top and side view of a $2\times2\times2$ lizardite supercell.
    (c) Top and side view of a $2\times2\times2$ brucite supercell.}
    \label{fig:primitive_cell}
\end{figure}

It should also be noted that the conventional unit cells of even $m$ antigorite can be further reduced to primitive cells with only half the number of atoms in the conventional cell setting, as shown in Fig.~\ref{fig:primitive_cell}. Fig.~\ref{fig:primitive_cell}(a) shows the top view of the $m=16$ antigorite supercells of the primitive cells containing a conventional cell.
The primitive cell, shown in Fig.~\ref{fig:primitive_cell}(b), preserves the symmetry of the monoclinic $C2/m$ space group.
Even though the symmetry in the primitive cell does not look as intuitive as the conventional cell, it dramatically increases the efficiency in the DFT calculations.

\subsection{Chemical environment analysis}
The structures of serpentines, shown in Fig.~\ref{fig:all_structures} and \ref{fig:primitive_cell}, including antigorite and lizardite, share many similarities. 
Some referred to lizardite as antigorite with $m = 1$ \cite{mookherjeeStructureElasticitySerpentine2009} while others as $m=\infty$ antigorite \cite{melliniAntigoritePolysomatismBehaviour1987}.
Another mineral, brucite, with layered Mg-O Octahedral and chemical formula of [Mg(OH)$_2$] also shares some of the structural features in the serpentines, including layered structure and inter-layer H-bonds [Fig.~\ref{fig:primitive_cell}(d)].

\begin{figure}[htbp]
    \centering
    \includegraphics[width=.5\textwidth]{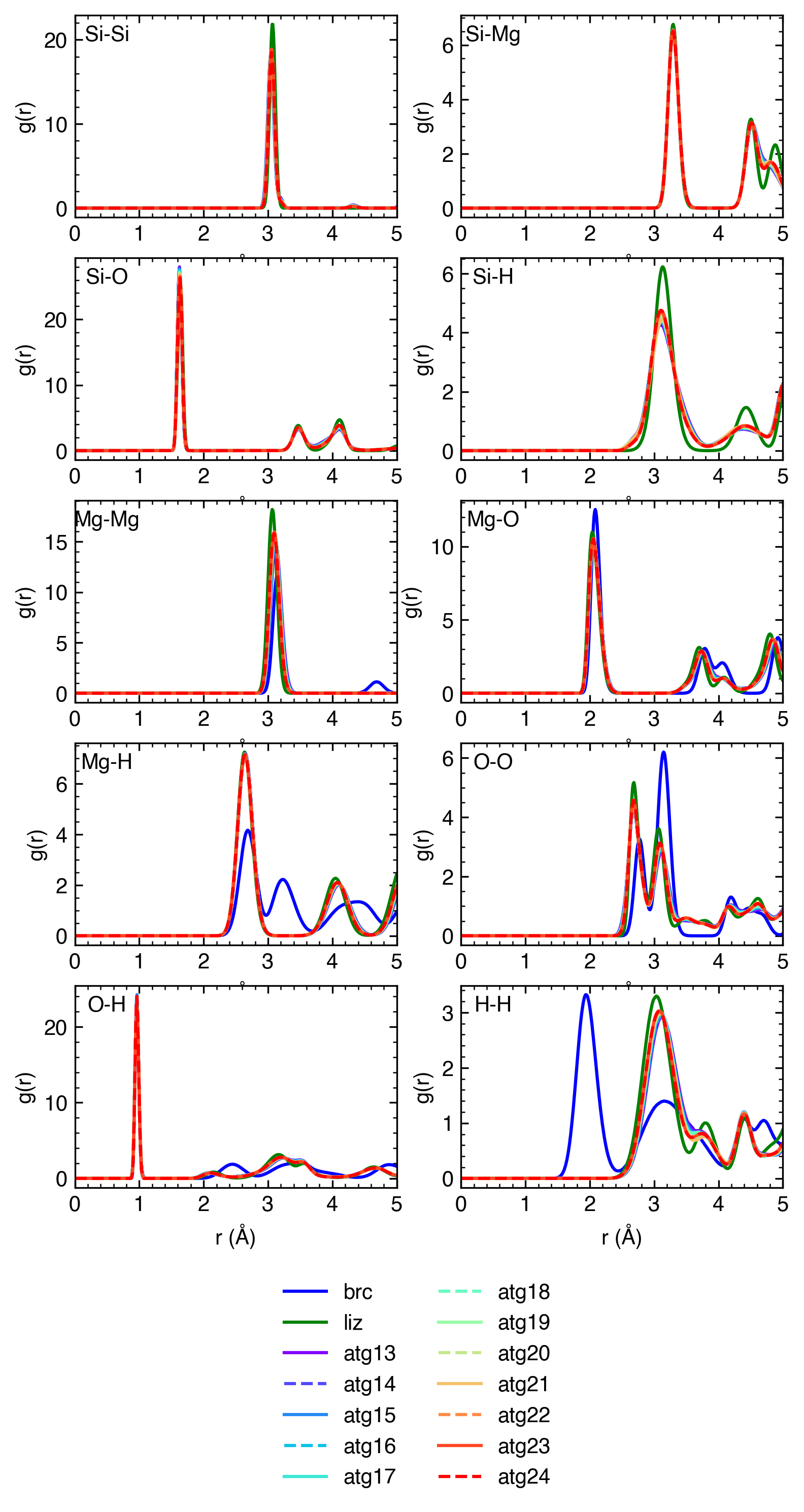}
    \caption{Partial pair correlation functions $g(r)$ for serpentine and brucite at 300~K and 0~GPa. The element pairs are listed in the upper left corner of each panel.}
    \label{fig:rdf_p}
\end{figure}

One natural question that arises here is, how similar/different are these structures?
Here, we analyze the partial pair correlation function of antigorite, lizardite, and brucite shown in Fig.~\ref{fig:rdf_p}.
Even though silicon is missing from the chemical formula of brucite, the chemical environment of magnesium is similar between brucite and serpentines, as shown in the Mg-Mg and Mg-O subplots.
However, the chemical environment of hydrogen is different between brucite and serpentines.
The chemical environment of antigorite with different wave periodicity and lizardite are very similar.
The chemical environment differences among the serpentine species are negligible compared to brucite.

\begin{figure}[htbp]
    \centering
    \includegraphics[width=.5\textwidth]{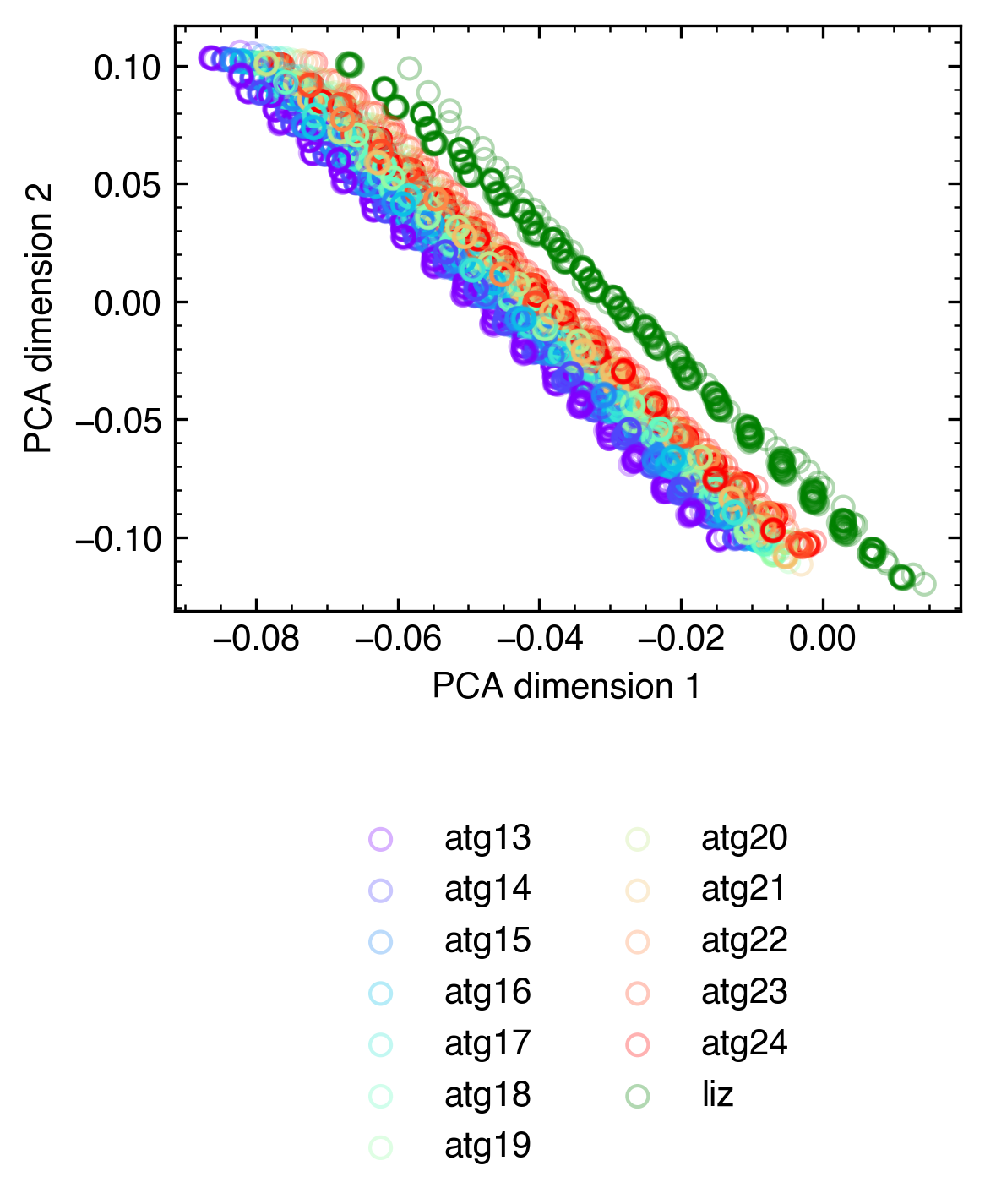}
    \caption{Principal component analysis (PCA) of the chemical environment feature matrix of serpentines. The chemical environment feature matrix has the size of $\#$ of configurations $\times$ $\#$ of features used to describe the chemical environment of the configuration. PCA is used to reduce the dimension of the matrix to 2. Each point in this figure represents a serpentine configuration. The distance between the points describes the similarity of the configurations: the closer the distance, the more similar the configurations are.}
    \label{fig:PCA}
\end{figure}

To further analyze the differences within the serpentine groups, we randomly sampled 7392 serpentine configurations from our r$^2$SCAN-DPMD simulations.
The chemical environment of these configurations is described using SOAP descriptors \cite{dscribe2}.
SOAP encodes regions of atomic geometries by using a local expansion of a Gaussian-smeared atomic density with orthonormal functions based on spherical harmonics and radial basis functions.
Using SOAP, we encode the chemical environments of the serpentine configurations into a matrix with the size of $\#$ of configurations $\times$ $\#$ of features to describe the configuration.
The number of features obtained using the setting described in the Methods section is 10583.
In this case, the dimension of the vector space represented by this matrix is 10583.
The high dimensional feature matrix is further reduced using PCA and plotted in Fig.~\ref{fig:PCA}.
The data of each serpentine species falls along a line, as our r$^2$SCAN-DPMD simulations are performed uniformly between 300--900~K and 0--15~GPa.
It can also be noted that with the increasing periodicity parameter $m$, the data shown in Fig.~\ref{fig:PCA} shift uniformly from left to right. 
The leftmost points are from antigorite with $m = 13$, and the rightmost points are from lizardite, with $m = 14 - 24$ in the middle.
This indicates that lizardite is indeed antigorite with $m = \infty$ instead of $m = 1$.

\subsection{Performance of machine learning potential in magnesium silicate hydrate (MSH) system}
\begin{figure}
    \centering
    \includegraphics[width=.5\textwidth]{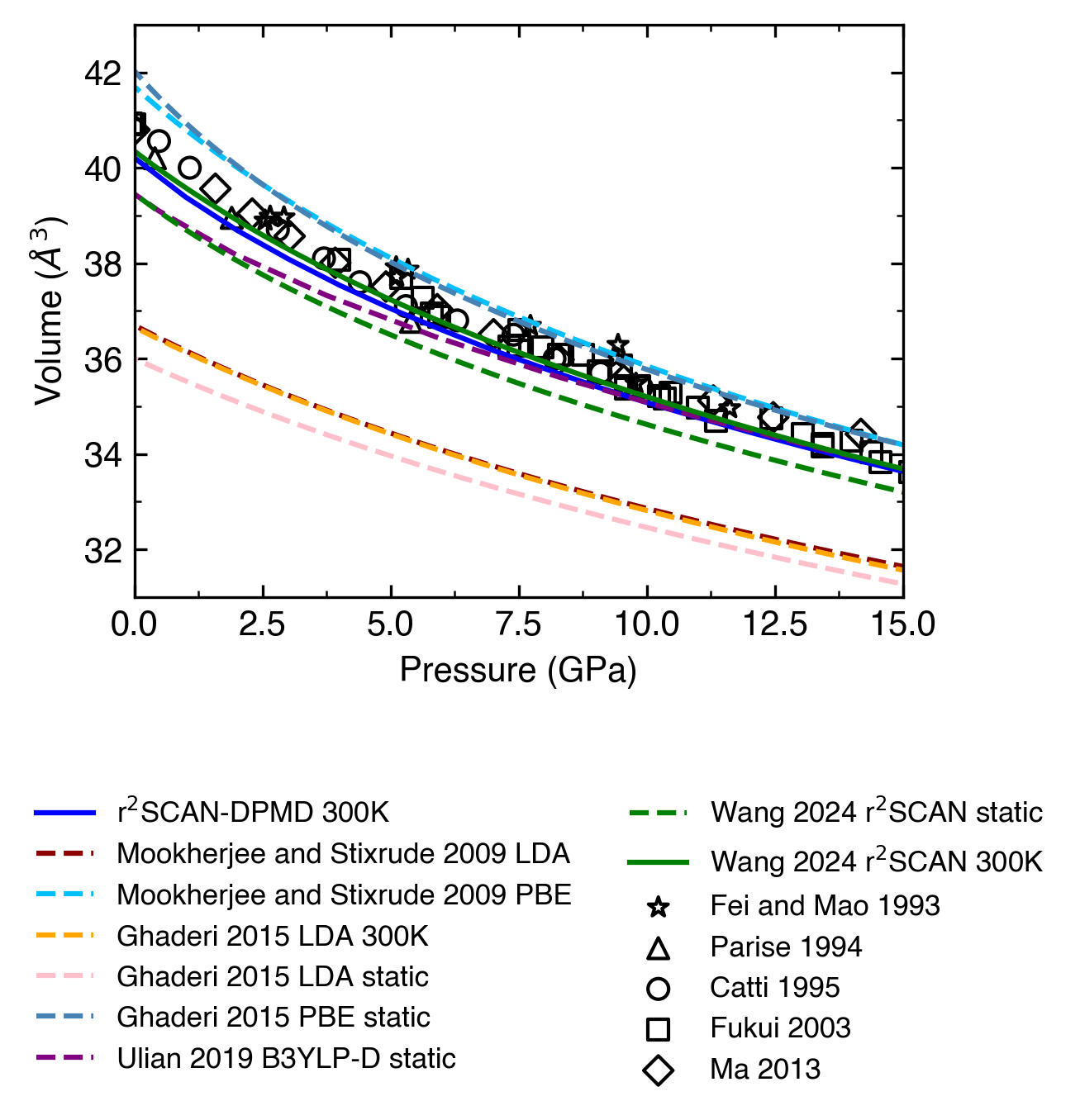}
    \caption{Equation of state of brucite.
    All measurements are marked with scattered symbols \cite{feiStaticCompressionMg1993,parisePressureinducedBondingNeutron1994,cattiStaticCompressionDisorder1995,fukuiThermalExpansionMg2003,maCompressionStructureBrucite2013}.
    All calculations are plotted using curves \cite{mookherjeeStructureElasticitySerpentine2009,ghaderiRelativeStabilityContrasting2015,dengElasticAnisotropyLizardite2022,wangInitioStudyStability2024}.}
    \label{fig:brc_eos}
\end{figure}

The high-temperature equation of state (EoS) is one of the most important properties in mineral physics.
Whether a theoretical method can reproduce the experimental EoS well is a gold standard for validating theoretical techniques.
In Fig.~\ref{fig:brc_eos}, we compare the 300~K EoS of brucite computed using r$^2$SCAN-DPMD with previous calculations \cite{mookherjeeHighpressureProtonDisorder2006,ghaderiRelativeStabilityContrasting2015,ulianEquationStateSecondorder2019,wangInitioStudyStability2024} and measurements \cite{feiStaticCompressionMg1993,parisePressureinducedBondingNeutron1994,cattiStaticCompressionDisorder1995,fukuiThermalExpansionMg2003,maCompressionStructureBrucite2013}.
Our 300~K r$^2$SCAN-DPMD reproduces the 300~K EoS calculated using r$^2$SCAN and quasiharmonic approximation (QHA) \cite{wangInitioStudyStability2024}, and among all calculations agree best with measurements.
This reemphasizes the success of r$^2$SCAN applied in the brucite and the ability of DPMD to reproduce the DFT accuracy.

As a simpler mineral in the serpentine family, lizardite is reported to undergo a phase transition around 5--6~GPa \cite{andreaniMicrostructuralStudyCrackseal2004,hilairetEquationsStateRelative2006a}.
Among all previous calculations \cite{mookherjeeStructureElasticitySerpentine2009,ghaderiRelativeStabilityContrasting2015,dengElasticAnisotropyLizardite2022},
only \citeA{tsuchiyaFirstprinciplesCalculationElastic2013,ghaderiRelativeStabilityContrasting2015} reported such transition, but at above 10~GPa and 8~GPa, respectively.
Our r$^2$SCAN-DPMD calculation is the only one that reproduces the EoS anomaly at $\sim$5--6~GPa, and it is also in better agreement with measurements than all previous calculations \cite{hilairetEquationsStateRelative2006a}.

\begin{figure}
    \centering
    \includegraphics[width=.5\textwidth]{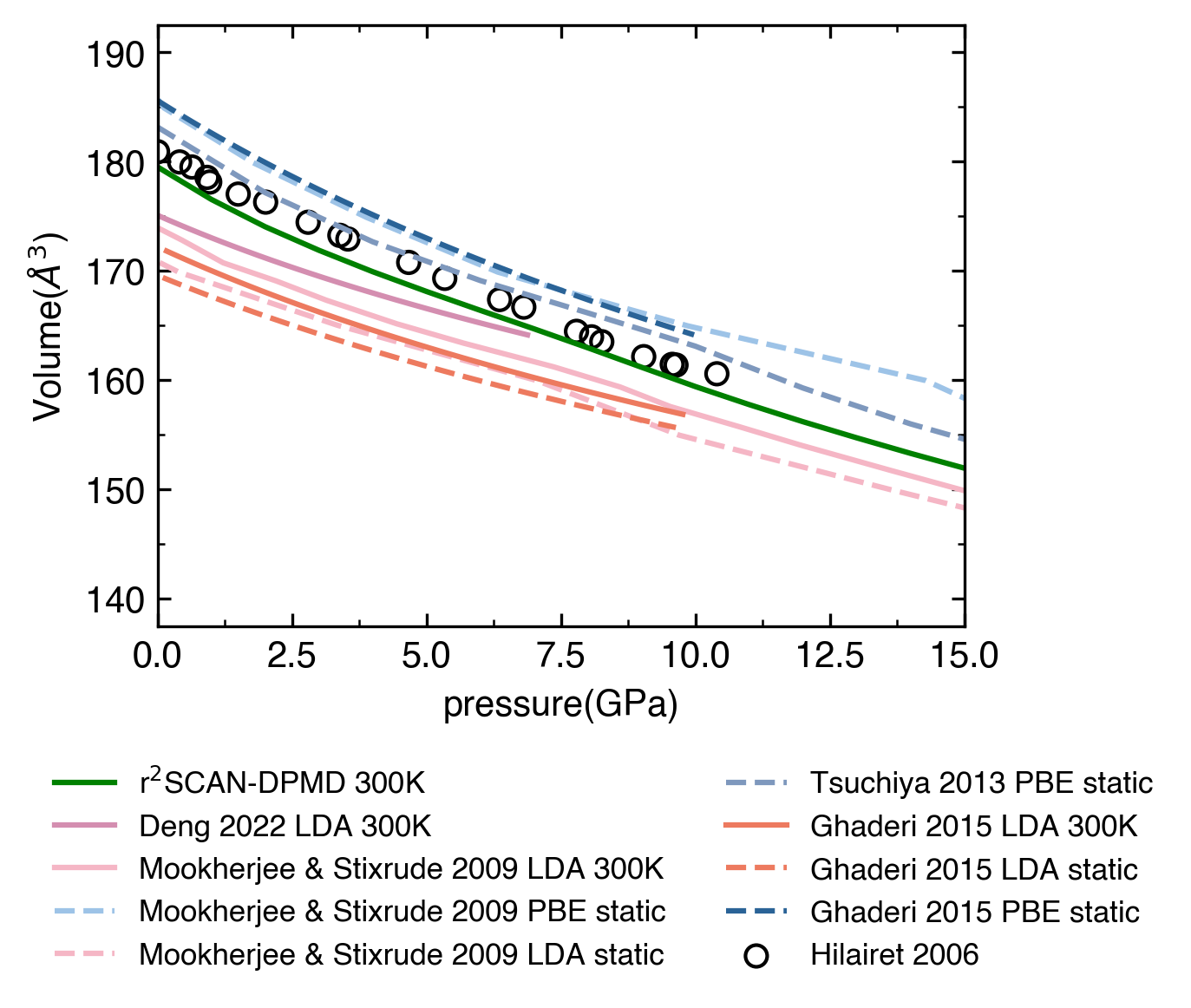}
    \caption{Equation of state of lizardite.
    Measurements \cite{hilairetEquationsStateRelative2006a} are marked with open circles.
    All calculations \cite{mookherjeeStructureElasticitySerpentine2009,ghaderiRelativeStabilityContrasting2015,dengElasticAnisotropyLizardite2022} are shown as colored curves.
    }
    \label{fig:liz_eos}
\end{figure}

\begin{figure}
    \centering
    \includegraphics[width=.5\textwidth]{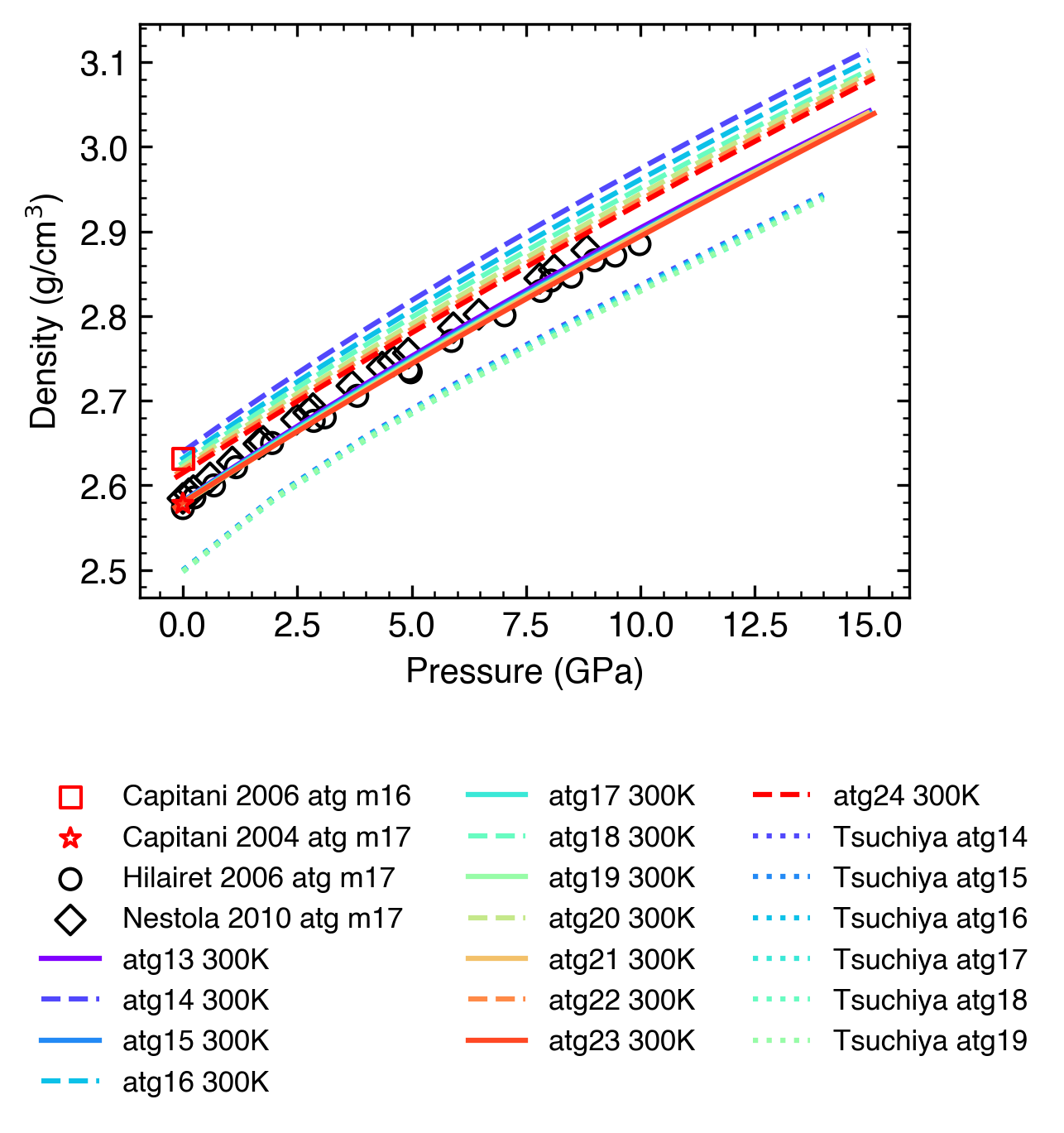}
    \caption{Equation of state of antigorite with different periodicity.
    All measurements are marked with scattered symbols \cite{capitaniModulatedCrystalStructure2004a,capitaniCrystalStructureSecond2006b,hilairetEquationsStateRelative2006a,nestolaAntigoriteEquationState2010}.
    All calculations are shown as colored curves \cite{tsuchiyaFirstPrinciplesInvestigations2024}.}
    \label{fig:atg_eos}
\end{figure}

In Fig.~\ref{fig:atg_eos}, we compare the 300~K EoS of antigorite calculated with r$^2$SCAN-DPMD and compared with previous measurements \cite{capitaniModulatedCrystalStructure2004a,capitaniCrystalStructureSecond2006b,hilairetEquationsStateRelative2006a,nestolaAntigoriteEquationState2010} and calculations \cite{tsuchiyaFirstPrinciplesInvestigations2024}. 
The EoS is plotted as density vs.\ pressure due to the polymorphism of antigorite with different periodicity.
Our results suggest that the EoS of antigorite with even and odd $m$ are quite different. 
This is expected as the antigorite with even $m$ has the $C2/m$ space group, whereas the antigorite with odd $m$ has the $Pm$ space group.
According to the geometrical model proposed by \citeA{grobetyPolytypesHigherorderStructures2003}, antigorite with even $m$, has a smaller amplitude of the wave than the odd ones, resulting in smaller volume and thus higher density.
Our results also show excellent agreement with previous measurements.
In contrast, recent similar results \cite{tsuchiyaFirstPrinciplesInvestigations2024} on antigorites do not display this distinction between $m$ even and odd EoSs.
This indicates that our r$^2$SCAN-DPMD results are more accurate.
The 300~K EoSs of brucite, lizardite, and antigorite are listed in the Supporting Information Table~S2. 

\begin{figure}
    \centering
    \includegraphics[width=.5\textwidth]{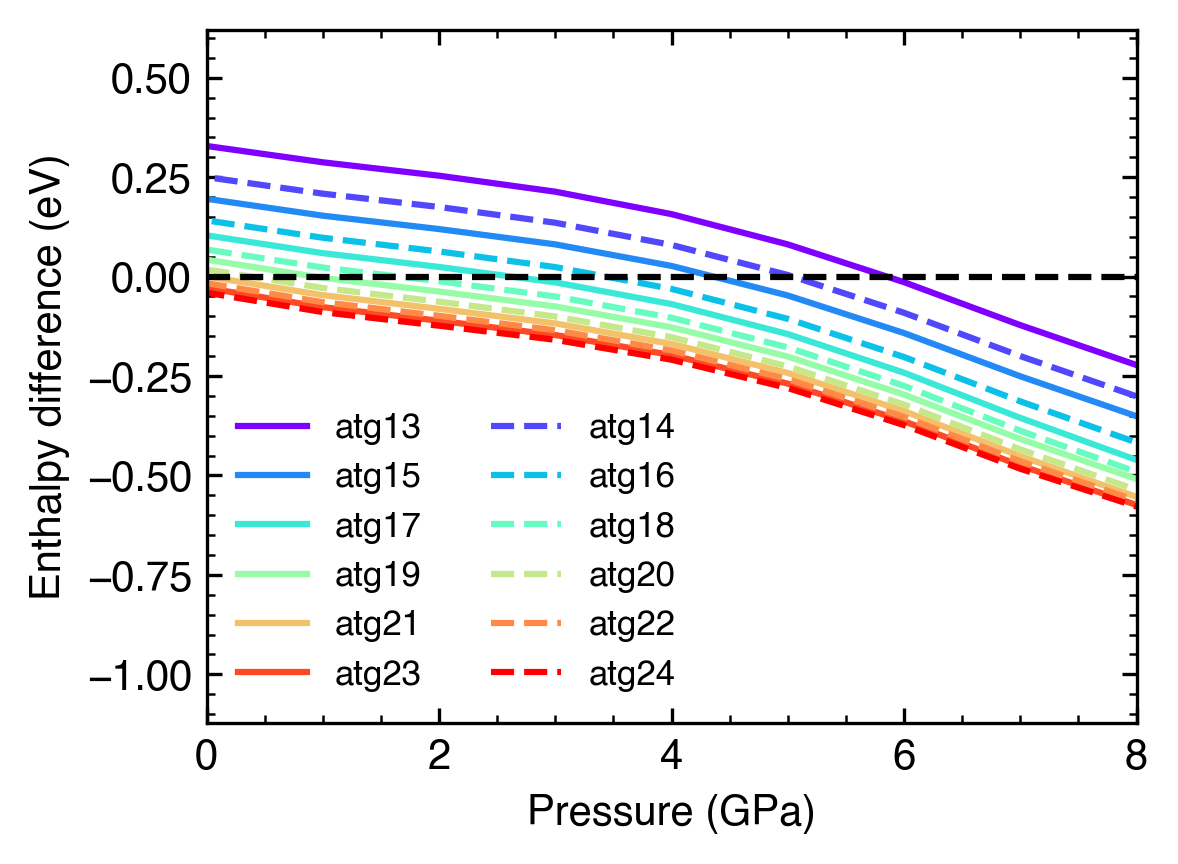}
    \caption{Enthaplpy difference, $\Delta H = \frac{3}{m} \, H_\mathrm{brucite} + \frac{1}{m} \, H_\mathrm{antigorite} - H_\mathrm{lizardite}$, vs.\ $P$.
    Negative (positive) $\Delta H$ corresponds to a more (less) stable brucite + antigorite combination than lizardite.
    }
    \label{fig:enthalpy_diff}
\end{figure}

With reliable static EoSs of brucite, lizardite, and antigorite, we further investigate the relative phase stability under the following chemical reaction:
\[
\mathrm{lizardite} \rightarrow \frac{3}{m} \, \mathrm{brucite} + \frac{1}{m} \, \mathrm{antigorite}
\]
In Fig.~\ref{fig:enthalpy_diff}, we show the enthalpy difference, $\Delta H$, of the above reaction for antigorite with all $m$s.
Negative (positive) $\Delta H$ corresponds to a more (less) stable brucite + antigorite combination than lizardite.
We notice that when $m$ increases, the relative stable range of lizardite decreases from around 6~GPa for antigorite with $m = 13$.
When $m > 21$, the antigorite + brucite is always more stable than lizardite. 
This might provide new insights into the low-temperature lizardite and antigorite relative stability as it is believed antigorite might be more stable than lizardite at very low temperatures ($<25^\circ $C) \cite{evansSerpentiniteMultisystemRevisited2004a}.

\section{Conclusion}
\label{sec:conclusion}
We developed a machine learning potential model for molecular dynamics simulations of serpentine minerals based on the r$^2$SCAN meta-GGA DFT functional.
The ML potential reproduces well the 300~K experimental EoSs of brucite, lizardite, and antigorite minerals at subduction zone conditions.
In addition, we investigate the polymorphism of antigorite with the periodicity $m$ = 13--24, which is believed to be all the naturally existent antigorite species.
When $m$ is larger than 21, the antigorite might be more stable than lizardite at low temperatures.
We believe the ML potential can be further applied to investigate the superstructure of antigorite with higher order when there is more than one wave in the structure.
This method may also be applied to other hydrous minerals in the Earth's interior.

\section*{Open Research Section}
The information on \textsc{DeePMD-kit}, the open‐source software is available at \url{https://github.com/deepmodeling/deepmd-kit}.
The information on VASP, first principles simulation software is available at \url{https://www.vasp.at}. 
All the necessary conditions for simulations are described in the Method section. 
The raw data of figures (Equation of states, enthalpy differences) shown in this manuscript can be obtained at (zenodo link, yet to be published).

\acknowledgments

This work was supported by DOE Award DE-SC0019759. Calculations were performed on the Extreme Science and Engineering Discovery Environment (XSEDE) \cite{townsXSEDEAcceleratingScientific2014} supported by the NSF grant \#1548562 and Advanced Cyberinfrastructure Coordination Ecosystem: Services \& Support (ACCESS) program, which is supported by NSF grants \#2138259, \#2138286, \#2138307, \#2137603, and \#2138296 through allocation TG-DMR180081. Specifically, it used the \textit{Bridges-2} system at the Pittsburgh Supercomputing Center (PSC), the \textit{Anvil} system at Purdue University, the \textit{Expanse} system at San Diego Supercomputing Center (SDSC), and the \textit{Delta} system at National Center for Supercomputing Applications (NCSA).

%
%

\bibliography{serpentine}

%
%
%
%
%

\end{document}


%
%


\title{Supporting Information for ``Machine learning potential for serpentines"}
%
%

%
%



\authors{Hongjin Wang\affil{1}, Chenxing Luo\affil{1}, Renata M.~Wentzcovitch\affil{1,2,3,4}}


\affiliation{1}{Department of Applied Physics and Applied Mathematics, Columbia University, New York, New York 10027, USA}
\affiliation{2}{Department of Earth and Environmental Sciences, Columbia University, New York, New York 10027, USA}
\affiliation{3}{Lamont--Doherty Earth Observatory, Columbia University, Palisades, New York 10964, USA}
\affiliation{4}{Data Science Institute, Columbia University, New York, New York 10027, USA}



%
%

%

\begin{article}

%
%

\noindent\textbf{Contents of this file}
\begin{enumerate}
\item Tables S1 to S2
\end{enumerate}
\end{article}
\clearpage

\begin{table}
\settablenum{S1}
\caption{Training and test dataset used to develop the machine learning potential in this paper. Numbers indicate the number of structures of each species.}
\centering
\begin{tabular}{c|c|c}
\hline
          & Training & Test \\
\hline
Brucite   & 1380  & 348  \\
Lizardite & 134   & 34   \\
m13       & 220   & 55   \\
m14       & 25    & 6    \\
m15       & 240   & 55   \\
m16       & 36    & 9    \\
m17       & 256   & 65   \\
m18       & 28    & 7    \\
m19       & 22    & 6    \\
m20       & 23    & 7    \\
m21       & 20    & 6    \\
m22       & 34    & 8    \\
m23       & 52    & 14   \\
m24       & 29    & 7    \\
\hline
Total     & 2499  & 631 \\
\hline
\end{tabular}
\end{table}

\begin{table}[]
\settablenum{S2}
\caption{300K EoS parameters of brucite, lizardite, and antigorite fitted using 3rd order Birch–Murnaghan EoS in this paper.}
\centering
\begin{tabular}{ccccc}
\hline
           & \multicolumn{1}{c}{$V_0$(\AA$^3$)}        & \multicolumn{1}{c}{$B_0$ (GPa)}       & \multicolumn{1}{c}{$B^\prime$}  & P-range (GPa)    \\
\hline
Brucite    & 40.19                         & 46.92                        & 6.73                        &  0--15  \\
\hline
Lizardite  & 179.42                        & 59.68                        & 7.65                        & 0--7  \\
           & 181.45 & 60.66 & 3.74 & 7--15  \\
\hline
Antigorite & \multicolumn{1}{c}{}          & \multicolumn{1}{c}{}         & \multicolumn{1}{c}{}        &          \\
m13        & 2,204.83                      & 70.22                        & 3.26                        &  0--15  \\
m14        & 4,770.70                      & 68.22                        & 3.47                        &  0--15  \\
m15        & 2,561.73                      & 71.69                        & 3.09                        &  0--15  \\
m16        & 5,484.62                      & 69.98                        & 3.28                        &  0--15  \\
m17        & 2,918.94                      & 72.33                        & 3.04                        &  0--15  \\
m18        & 6,196.74                      & 71.13                        & 3.24                        &  0--15  \\
m19        & 3,277.89                      & 72.57                        & 2.87                        &  0--15  \\
m20        & 6,909.35                      & 73.61                        & 2.82                        &  0--15  \\
m21        & 3,634.90                      & 73.46                        & 2.74                        &  0--15  \\
m22        & 7,621.48                      & 75.11                        & 2.65                        &  0--15  \\
m23        & 3,989.86                      & 74.99                        & 2.67                        &  0--15  \\
m24        & 8,336.09                      & 75.40                        & 2.61                        &  0--15
\\
\hline
\end{tabular}
\end{table}